\documentclass[smallcondensed]{svjour3}

\usepackage[utf8]{inputenc}

\usepackage{babel}
\usepackage{hyperref}
\usepackage{graphicx,xcolor}
\usepackage{subcaption}
\usepackage{amssymb,amsmath}
\usepackage{enumerate}
\usepackage{booktabs}
\usepackage[linesnumbered,lined,commentsnumbered]{algorithm2e}
\usepackage{setspace}
\usepackage{tikz}

\DeclareMathOperator{\im}{Im}     

\begin{document}

\title{Geometric goodness of fit measure to detect patterns in data point clouds}

\author{Alberto J. Hern\'andez \and Maikol Sol\'is \and Ronald A.
	Z\'u\~niga-Rojas}

\institute{Alberto J. Hern\'andez \at Centro de Investigaci\'on en Matem\'atica
	Pura y Aplicada (CIMPA), Escuela de Matem\'atica, Universidad de Costa Rica,
	San Jos\'e, Costa Rica \email{albertojose.hernandez@ucr.ac.cr}\and %
	Maikol Sol\'is (Corresponding author) \at Centro de Investigaci\'on en Matem\'atica Pura y Aplicada
	(CIMPA), Escuela de Matem\'atica, Universidad de Costa Rica, San Jos\'e, Costa
	Rica \email{maikol.solis@ucr.ac.cr}\and %
	Ronald A. Z\'u\~niga-Rojas \at Centro de Investigaci\'on Matemáticas y
	Meta-Matem\'aticas (CIMM), Escuela de Matem\'atica, Universidad de Costa Rica,
	San Jos\'e, Costa Rica \email{ronald.zunigarojas@ucr.ac.cr} }
\maketitle

\begin{abstract}
We derived a geometric goodness-of-fit index, similar to \(R^{2}\)   using topological data analysis techniques. We build the Vietoris-Rips complex from the  data-cloud projected onto each variable. Estimating the area of the complex and their domain,    we create an index that measures the emptiness of the space with respect to the data. We made the analysis with an own package    called TopSA (Topological Sensitivy Analysis).



	%
	%
\end{abstract}



\section*{Introduction}
\label{sec:0}

Data point cloud recognition is an essential task in any statistical procedure. Finding a pattern into the data expose the inherent characteristics of the phenomena we want to model. There are tools described in the literature  like linear regression or clustering  used to achieve this \cite{Hastie2009,Hand2005}. Other research branches use data visualization techniques to highlight features hidden in the data \cite{Tufte2001,Myatt2009,Buja2005}.

Professionals in computational modeling aim to reduce their models by choosing their problems’ most relevant factors. One way to achieve this is through goodness-of-fit measures, used to find the relevance of certain chosen models to explain a variable.  One classic way to determine if  some variables fit inside a model is using the determination coefficient \(R^2\). This quantity measures the amount of variance explained by some model against the variance explained by the model formed only by a constant.  It measures how preferable it would be to fit a
model against just fitting  a  constant.  If \(R^2\) is near  one, then the model fits well into the data. Otherwise it is better to adjust to a constant.

The \(R^2\) method  has been controversial since its origins \cite{Barrett1974}. For example, we can increase the \(R^2\) score by adding new variables to the model. Even if they are unnecessary to the problem the \(R^2\) will increase. As a general rule, high \(R^2\) does not imply causation between covariance and outcome.

Some authors have proposed some extensions and properties of this score throughout the years. The work of~\cite{Barten1962} presents a bias-reduced \(R^2\). In \cite{Press1978} the authors conducted a Bayesian analysis. In ~\cite{Cramer1987} the authors showed that in small samples, the \(R^2\) score is higher. Even with those constraints, \cite{Barrett1974} concludes how useful are these measures in applied research.

This goodness-of-fit measures overlook the geometric arrangement of the data. They build the statistical relation between $X$ and $Y$ and then present it in the form of an indicator. Depending on this simplification, they do not consider the geometric properties of the data. For example, most indices will cannot recognize geometric structure when the input variable is zero-sum, treating it as random noise.

The study of topological data is a recent field of research that aims to overcome these shortcomings. Given a set of points generated in space, it tries to reconstruct the model through an embedded manifold that covers the data set. With this manifold, we can study the characteristics of the model using topological and geometrical tools instead of using classic statistical tools.

Two classical methods used to describe the intrinsic geometry of the data are Principal Components Analysis (PCA) and Multidimensional Scaling (MDS). PCA transforms the data into a smaller linear space, preserving the statistical variance. MDS performs the same task but preserves the distances between points. Methods like the isomap algorithm developed in~\cite{Tenenbaum2000} and expanded in~\cite{Bernstein2000} and \cite{Balasubramanian2002} unify these two concepts to allow the reconstruction of a low-dimensional variety for non-linear functions. Using geodesic distance, isomap identifies the corresponding manifold and searches lower dimensional spaces.

In recent years new theoretical developments have used theorical ideas in topology for the data analysis. Such  ideas include persistent homology, simplicial complexes, and Betti numbers. Reconstructing manifolds work for clouds of random data and functional data. The following works present some examples: \cite{Ghrist2008},~\cite{Carlsson2009}  and \cite{Carlsson2014}. In~\cite{Gallon2013} and in~\cite{Dimeglio2014}. Also, this approach compacts \textit{``Big Data’’} problems efficiently  (see ~\cite{Snasel2017}).

In this work we aim to connect the concept of goodness of fit with the analysis of topological data  through a geometrical \(R^2\) index. By doing this it will be possible to determine what variables have structured patterns using the geometrical information extracted from the data.

The outline of this study is: Section~\ref{sec:preliminary} deals with key notions, both in sensitivity analysis and in topology. In Subsection~\ref{sec:goodness} wer review and comment classic methods to determine the $R^2$. We finish this subsection with an example which motivated the work in this paper. Subsection~\ref{sec:VR-complex} deals with preliminaries in homology and describes the Vietoris-Rips Complex. Section~\ref{sec:methodology} explains the method used to create our sensitivity index; Subsection~\ref{sec:neigborhood-graph} constructs the neighborhood graph, and deals with different topics such as \textit{the importance of scale}, the \textit{Ishigami Model} and presents programming code to determine the radius of proximity. Subsection~\ref{sec:VR-expansion} describes the algorithm used to construct the homology complex through the \textit{Vietoris-Rips  complex} and Subsection~\ref{sec:goodness-index} explains our proposed sensitivity index. Section~\ref{sec:results} describes our results, it describes the software and packages used to run our theoretical examples. Subsection~\ref{sec:theoretical-examples} is a full description of each theoretical example together with visual aids, such as graphics and tables describing the results. Section~\ref{sec:conclusions} contains our conclusions and explores scenarios for future research. 

\section{Preliminary Aspects}\label{sec:preliminary}

In this section we will discuss the context and tools needed to implement our geometric goodness-of-fit.

\subsection{Measuring  goodness-of-fit}\label{sec:goodness}

Let $(X_1, X_2, \ldots X_{p}) \in \mathbb {R} ^ {p}$ for $p \geq 1$ and $Y \in
	\mathbb {R}$ two random variables. Define the non-linear regression model as
\begin{equation}
	\label{eq:regression_nonlinear}
	Y = \varphi (X_1, X_2, \ldots X_{p}) + \varepsilon,
\end{equation}
where $\varepsilon$ is random noise independent of $(X_1, X_2, \ldots X_{p})$. The
unknown function $\varphi: \mathbb {R} ^ {p} \mapsto \mathbb {R}$ describes the
conditional expectation of $Y$ given $(X_1, X_2, \ldots X_{p})$. Suppose as well
that $(X_{i1}, X_{i2} \ldots X_{ip}, Y_i)$, for $i = 1, \ldots, n$, is a size $n$
sample for the random vector$(X_{1}, X_{2}\ldots X_{p}, Y) $.

If $p\gg n$ the model~\eqref{eq:regression_nonlinear} suffers from the
\emph{``curse of dimensionality''}, term introduced in~\cite{Bellman1957} and
~\cite{Bellman1961}, where is shown that the sample size $n$, required to fit a model
increases with the number of variables $p$. Model selection techniques solve
this problem using indicators as the AIC or BIC, or more advanced techniques
such as Ridge or Lasso regression. For the interested reader there is a
comprehensive survey in~\cite{Hastie2009}.

Suppose in the context
of the linear  regression we have the model
\begin{equation*}
	\mathbf{Y} =    \mathbf{X} \mathbf{b} + \boldsymbol{\varepsilon}
\end{equation*}

where

\begin{equation*}
	\mathbf{Y} = \begin{pmatrix}
		Y_1    \\
		\vdots \\
		Y_n
	\end{pmatrix},
	\
	\mathbf{X} = \begin{pmatrix}
		1      & X_{11} & X_{12} & \cdots & X_{1p} \\
		\vdots &        & \vdots &        & \vdots \\
		1      & X_{n1} & X_{n2} & \cdots & X_{np} \\
	\end{pmatrix},
	\
	\mathbf{b} =   \begin{pmatrix}
		b_0    \\
		\vdots \\
		b_p
	\end{pmatrix},
	\
	\boldsymbol{\varepsilon} = \begin{pmatrix}
		\varepsilon_1 \\
		\vdots        \\
		\varepsilon_n
	\end{pmatrix}
\end{equation*}

and \(\boldsymbol{\varepsilon}\) is a noisy vector with mean
\((0,\ldots,0)^\top\) and identity covariance.

The least-square solutions to find the best coefficients is

\begin{equation*}
	\mathbf{\hat{b}}  = \left(  \mathbf{X} ^\top  \mathbf{X} \right)^{-1}
	\mathbf{X} ^\top   \mathbf{Y}.
\end{equation*}

If \(p = 1\) the problem reduces to the equations,
\begin{align*}
	\hat{b}_{i1} & = \frac{\sum_{j=1}^n \left(X_{ij} - \bar{X}_i  \right)
		\left( Y_j - \bar{Y} \right)} {\sum_{j=1}^n \left( X_{ij}
		-X_i \right)^2 }  \\
	\hat{b}_{i0} & = \bar{Y} - \hat{b}_1 \bar{X}_i
\end{align*}

Notices that in the particular case \(p = 0\) (the null
model) then the estimated parameter simplifies into $\hat{b}_{0i}=\bar{Y}$.

The following identity holds in our context,

\begin{align*}
	\sum_{j=1}^n \left(Y_j  - \bar{Y}  \right)^2 & =
	\sum_{j=1}^n \left(\hat{Y}_j  - \bar{Y}  \right)^2
	+ \sum_{j=1}^n \left(\hat{Y}_j  -Y_j \right)^2
\end{align*}

One of the most used quantities to quantify if one covariate (or a set
of them) are useful to explain an output variable is the statistic
$R^2\in[0,1]$. We estimate it as
\begin{align*}
	R^2  = 1 - \frac{\sum_{j=1}^n \left(\hat{Y}_j  -Y_j \right)^2
	}{\sum_{j=1}^n \left(Y_j  - \bar{Y}  \right)^2 }
\end{align*}

This value indicates how much the variability of the regression model
explains the variability of \(Y\). If \(R^2\) is close to zero, the
squared residuals of the fitted variables  are similar to the residuals of a null
model formed only by a constant. Otherwise, the residuals of the null
model are greater than the residuals of the fitted values, meaning that
the selected model could approximate better the observed values of the
sample.

The $R^2$ has the deficiency that it increases if we add new variables
to the model. A better statistic to measure the goodness of fit but
penalizing the inclusion of nuisance variable is the Adjusted $R^2$,

\begin{equation*}
	R^2_{Adj} = 1- (1-R^2) \frac{n-1}{n-p-1}
\end{equation*}

These measures could detect if a data point cloud could be fitted
through some function. However, if the structure of the data has
anomalous  patterns $R^2$ could be insufficient.

\begin{figure}[htpb]
	\centering
	\includegraphics[width=\textwidth]{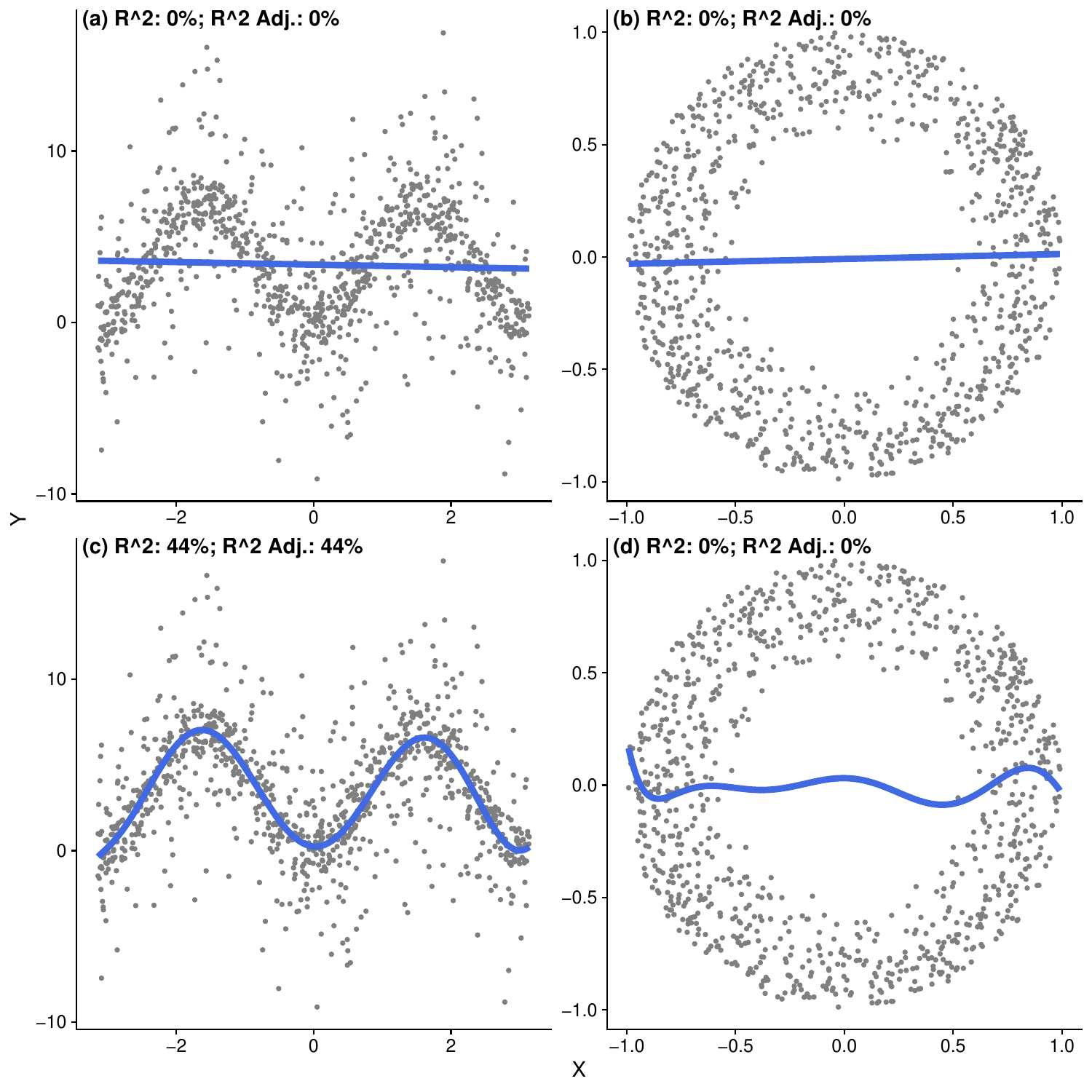}
	\caption{\emph{Linear regression  $Y=a_0 + a_1 X$} with \textbf{(a)}
		Ishigami model, and \textbf{(b)}  Circle with hole model.
		\emph{Polynomial regression $Y=a_0 + \sum_{k=1}^8 a_k X_1^k$} with
		\textbf{(c):} Ishigami model \textbf{(d):} Circle with one hole
		model.} \label{fig:fittedR2}
\end{figure}

For example  Figure~\ref{fig:fittedR2} presents this phenomena for two data sets. We adjusted a
polynomial of degree 8 to each set using least-square regression; the ``Ishigami'' and ``Circle with one hole'' models  are presented in Section~\ref{sec:results}.

The Ishigami model presents strong non-linearity in its second
variable. The model could capture the relevant pattern of the data and
we got a \(R^2 = 0.448 \) and \(R^2_{Adj} = 0.4435\) (panel
\textbf{(c)}). This tells us that we could capture around the 44\% of the
total variability of the data. We point out that classic regression failed to capture any of geometric features. In
particular the \(R^2\) and $R^2_{Adj}$ are near to zero.

The issue with  panels \textbf{(a)}, \textbf{(b)} and \textbf{(d)} in Figure~\ref{fig:fittedR2} is that the fitted models are inflexible with
respect to the data that we used. In particular, the ``Circle with one hole''
model requires a better understanding about the anomalous geometry of
the data cloud. The next Section explores how to better determine the geometric structure of the data to overcome this shortcomings.

\subsection{Homology and Vietoris-Rips Complex}\label{sec:VR-complex}

The seminal work of \cite{Carlsson2014} presents the ground basic definitions to define a homological structure for a data cloud of points.
In particular, we are interested in the reconstruction  of the embedding manifold of the cloud of points using only the distance between each pair of
points.

For the purpose of this paper a geometric object is either a connected surface or connected directed graph. Given a geometric object define a
$0$-simplex  as a point, frequently called a {\em vertex}. Since we deal with finite sets of data, taking coordinates on the Euclidean Plane
$\mathcal{E} = \mathbb{R}^2$, we denote a $0$-simplex as a point $p_j = (x_j, y_j)$ for $j = 1, \ \dots,\ n$.

If we join two distinct $0$-simplices, $p_0, p_1$, by an oriented line segment, we get a $1$-simplex called an {\em edge}:
$\overline{p_{0}p_{1}} = (p_{1}-p_{0})$.

Consider now three non-colinear points $p_0,\ p_1,\ p_2$ as $0$-simplices, together they form three $1$-simplices:
$\overline{p_0 p_1}$, $\overline{p_0 p_2}$ and $\overline{p_1 p_2} = \overline{p_0 p_2} - \overline{p_0 p_1}$.
This last equation shows that only two of them are linearly independent and span the other. The union of these three
edges form a triangular shape, a $2$-simplex called a {\em face}, denoted as $\triangle(p_0 p_1 p_2)$ that contains all the points
enclosed between the edges:

\[
	\triangle(p_0 p_1 p_2) =
	\Big\{
	p\in \mathbb{R}^2\colon
	p = \sum_{j=0}^{2}\lambda_j p_j \colon
	\sum_{j=0}^{2}\lambda_j = 1, \lambda_j\geq 0
	\Big\}.
\]

As well as $2$-simplices, if we consider four non-coplanar points on the
Euclidean Space $\mathbb{R}^3$, we can construct a $3$-simplex called
\emph{tetrahedron}. A generalization of dimension $n$ will be a convex set
in $\mathbb{R}^n$ containing $\left\{ p_0, p_1, \dots, p_n \right\}$ a subset of $n+1$ distinct
points that do not lie in the same hyperplane of dimension $n-1$ or,
equivalently, that the vectors $\left\{ \overline{p_0 p_j} = p_j - p_0\right\}, 0 <
	j\leq n $ are linearly independent. In such a case, we are denoting the points
$\left\{ p_0, p_1, \dots, p_n \right\}$ as vertices, and the usual notation would be
$[p_0, p_j]$ for edges,
$[p_0, p_1, p_2]$ for faces, $[p_0, \dots, p_4]$ for tetrahedra, and $[p_0,
			\dots, p_n]$ for $n$\emph{-simplices}. Nevertheles, for our purposes, it will
be enough to consider just $0,\ 1$ and $2$-simplices, the latter constituting
the  building blocks of the Vietoris-Rips complex.

A $\bigtriangleup$-complex is the topological quotient space of a collection of disjoint simplices identifying some of their faces. This complex will be ordered in the sense that each vertex  is in lexicographic order. In other words, there exists an injective mapping between \(\left\{ p_{0},\ldots, p_{n} \right\}\) and \(\mathbb{N}\). 

For any \(n-\)simplex  $\sigma = [p_0,\dots,p_n]$  the boundary of \(\sigma\) is the the linear combination of $(n-1)$-simplices of the form 
\begin{equation} 
\partial_{n} \sigma = \sum_{j=0}^n (-1)^j [p_0,\dots,\hat{p}_j,\dots,
p_n]\label{chain003}
\end{equation} 
where 
\begin{equation*}
[p_0,\dots,\hat{p}_j,\dots, p_n] = [p_0,\dots,p_{j-1},p_{j+1},\dots,
p_n].
\end{equation*}

Hence, give our oriented simplicial complex, of dimension \(n\), we can build a sequence of vector spaces and linear maps of the form, 

\begin{equation*}
\dots \to
\bigtriangleup_n(X)\xrightarrow{\partial_n}
\bigtriangleup_{n-1}(X)\xrightarrow{\partial_{n-1}}
\bigtriangleup_{n-2}(X)\to
\dots \to
\bigtriangleup_1(X)\xrightarrow{\partial_1}
\bigtriangleup_{0}(X) \xrightarrow{\partial_0}  0
\end{equation*}
where $\partial \circ\partial = \partial_{n} \circ\partial_{n-1} = 0$ for all $n$.
This is usually known as a \emph{chain complex}. Since $\partial_{n} \circ
\partial_{n-1} = 0$, the $\im(\partial_n) \subseteq \ker(\partial_{n-1})$, and so, we
define the $n^{th}$ \emph{simplicial homology group} of $X$ as the quotient
\begin{equation}
\label{eq:def-homology-group}
H_n^{\bigtriangleup}(X) = \frac{\ker(\partial_n)}{\im(\partial_{n+1})}.
\end{equation}
The elements of the kernel are known as \emph{cycles} and the elements of the
image are known as \emph{boundaries}.

The
$n^{th}$\emph{-Betti number} of $X$ is the rank of each space \(H_{n}^\bigtriangleup(X)\). Roughly speaking, the $n$-th Betti number represents the number of $n$-dimensional holes on the topological surface. Formally, from the algebraic topology stand point of view, the first Betti numbers is $b_0$, that represents the number of connected components of the manifold; the second one is $b_1$, that represents the number of loops (circles) that span $H_1$ or informally, the number of circular holes ($1$-dimensional holes) on the manifold, and $b_2$ that represents the number of {\em voids}, $2$-dimensional holes. The reader may see \cite{Hatcher2000} or \cite{book:72583}
for details.

Clarified those concepts, we can now define what is Vietoris-Rips complex. 

\begin{definition}[VR neigboorhood graph]
	\label{def:VR-neigborhood-graph} Given a sequence of vertices
\(\{p_0,\ldots,p_{n}\}\) \(\subseteq\) \(\mathbb{R}^{n}\)  and some
positive value $\varepsilon\in \mathbb{R}$, the VR neighborhood graph
is defined as $G_{\varepsilon}(V) = (V, E_{\varepsilon}(V))$ where
	\begin{align*}
	\label{eq:VR-neigborhood}
	V &= \{p_0,\ldots,p_{n}\} \\
	E_{\varepsilon}(V)  &= \{\{u,v\}\ \mid\ d(u,v)\leq\varepsilon, u\neq v \in V\},
	\end{align*}
	represent the vertices and edges respectively. 
\end{definition}

\begin{definition}[VR expansion]
	\label{def:VR-expansion}
	Given a neighborhood graph $G_{\varepsilon}(V)$, their Vietoris-Rips
	complex $\mathcal{R}_{\varepsilon}$ is defined as all the edges of a simplex $\sigma$ that
	are in $G_{\varepsilon}(V)$. In this case $\sigma$ belongs to $\mathcal{R}_{\varepsilon}$. For $G_{\varepsilon}(V) = (V, E_{\varepsilon}(V))$, we have
	\begin{equation*}
	\mathcal{R}_{\varepsilon} = V \cup E_{\varepsilon}(V) \cup \left\{\sigma \mid {\binom{\sigma}{2}}
	\subseteq E_{\varepsilon}(V)\right\}.
	\end{equation*}
	where $\sigma$ is a simplex of $G$.
\end{definition}

\subsection{Persistent Homology}

There are two classical mistakes that can be made when start working with homological complexes. The first one, trying to find the optimal value for the radio $\varepsilon$, the homology of the complex associated to a point cloud data at a specific radio will be not enough. The second one, Betti numbers will be insufficient: we need a tool that tells us which holes we can keep and which holes we can ignore; classic homological constructions do not have that flexibility.

Persistence is a solution to this problem. The concept of {\em persistence} was introduced by Carlsson~\cite{Carlsson2009}, Carlsson and Zomorodian~\cite{Zomorodian2005}, and Zomorodian~\cite{Zomorodian2010}, and also presented by Ghrist~\cite{Ghrist2008}. Consider a parametrized family of spaces, for instance a sequence of Rips complexes $\{\mathcal{R}_j\}_{j=1}^{n}$ associated to a specific point cloud data for an incresaing sequence of radii $(\varepsilon_j)_{j=1}^{n}$. Instead of considering the individual homology of the complex $\mathcal{R}_j$, consider the homology of the inclusion maps
\[
 \mathcal{R}_1
 \xrightarrow{i}
 \mathcal{R}_2
 \xrightarrow{i}
 \dots 
 \mathcal{R}_{n-1}
 \xrightarrow{i}
 \mathcal{R}_n
\]
{\em i.e.}, consider he homology of the iterated induced inclusions
\[
 i_{*}\colon 
 H_{*}(\mathcal{R}_j)
 \to
 H_{*}(\mathcal{R}_k)
 \quad 
 \textmd{for}
 \quad 
 j < k.
\]
These induced homology maps will tell us which topological features persist. The persistence concept tell us how Rips complexes become a good approximation to \u{C}ech complexes (For full details see \cite{DeSilva2007}, \cite{Zomorodian2005}).


\begin{lemma}[Lemma~2.1.\cite{Ghrist2008}]
 For any radio $\varepsilon > 0$ there are inclusions
 \[
  \mathcal{R}_{\varepsilon}
  \hookrightarrow
  \mathcal{C}_{\varepsilon \sqrt{2}}
  \hookrightarrow
  \mathcal{R}_{\varepsilon \sqrt{2}}.
 \]
\end{lemma}

To work with {\em persistent homology} in general, start considering a {\bf persistence complex} $\{C_{*}^{i}\}_{i}$ which is a sequence of 
chain complexes joint with chain inclusion maps $f^{i}\colon C_{*}^{i}\to C_{*}^{i+1}$, according to the tools we are working with, we can 
take a sequence of Rips or \u{C}ech complexes of increasing radii $(\varepsilon_j)_j$.

\begin{definition}
 For $j < k$, we define the $(j,k)$-persistent homology of the persistent complex $C = \{C_{*}^{i}\}_{i}$ as the image of the induced homomorphism 
 $f_{*}\colon H_{*}(C_{*}^{j})\to H_{*}(C_{*}^{k})$. We denote it by $H^{j\to k}_{*}(C)$.
\end{definition}

A final ingredient will be needed. The {\em barcodes} are graphical representations of the persistent homology of a complex $H^{j\to k}_{*}(C)$ as 
horizontal lines in a $XY$-plane whose corresponds to the increasing radii $\{\varepsilon_j\}_j$, and whose vertical $Y$-axis 
corresponds to an ordering of homology generators. Roughly speaking, a barcode is the persistent topology analogue to a ``Betti number''. 

\begin{figure}
	\centering
	\includegraphics[width=\linewidth]{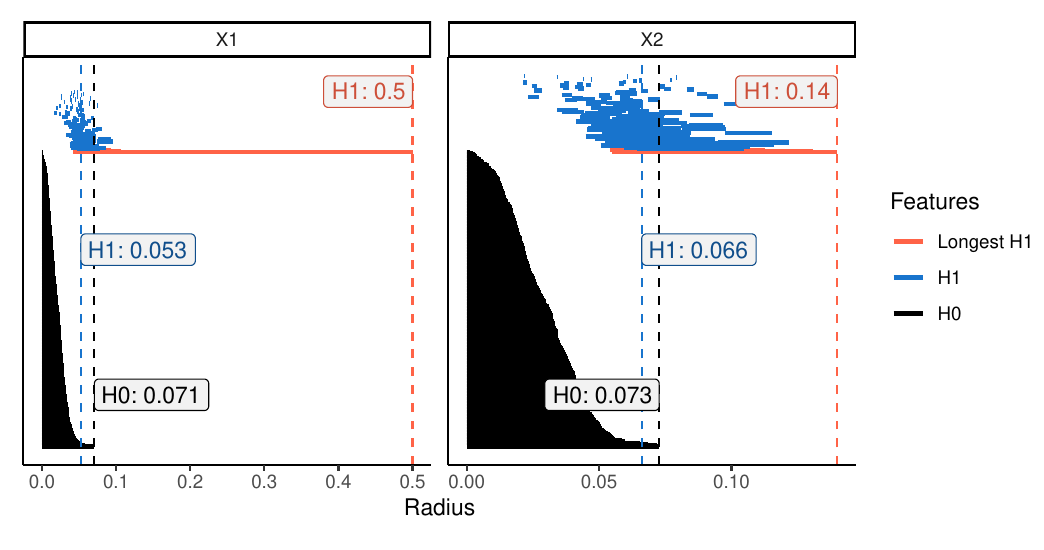}
	\caption{Barcode representing the topological features of Figure~\~ref{fig:fittedR2}}
	\label{fig:circ1-ex-barcode-crop}
\end{figure}

Figure~\ref{fig:circ1-ex-barcode-crop} represent the barcodes for the example of a circle with a hole from Figure~\ref{fig:fittedR2}. Notices how the central hole persist on the left side plot (long \(H_{1}\) red lines)  while the noise it is ~characterized by short lines in \(H_{1}\).

\section{Methodology}\label{sec:methodology}

Recall the model~\eqref{eq:regression_nonlinear}. The random variables
$(X_{1}, \ldots, X_{p})$ are distorted by the function $m$ and its
topology. Our aim is to measure how much each of the $X_{i}$ influence
this distortion, i.e.\ we want to determine which variables influence
the model the most.

In Section~\ref{sec:goodness} we reviewed  classical  statistics methods to
estimate the correspondence between random variables $X_{i}$, for $i = 1,\ldots, p$,
with respect to a model $Y$. In this paper we  consider the geometry
of the point-cloud, its enveloping manifold, and create an index that  reveals information about the model.

The first step is to create a neighborhood graph for the point-cloud
formed by \((X_{i}, Y)\) where an edge is set if a pair of nodes are
within \(\varepsilon\) euclidean distance. In this way we connect only
the nodes nearby within a fixed distance. With this neighborhood graph
we construct the persistent homology using the method
of~\cite{Zomorodian2010} for the Vietoris-Rips (VR) complex.

The algorithm of~\cite{Zomorodian2010} works in two-phases: First it
creates a VR neighborhood graph
(Definition~\ref{def:VR-neigborhood-graph}) and then builds the VR
complex step-by-step (Definition~\ref{def:VR-expansion}).

The definitions imply a two-phase procedure to construct the
Vietoris-Rips complex with resolution $\varepsilon$:
\begin{enumerate}
	\item Using Definition~\ref{def:VR-neigborhood-graph} compute the
	      neighborhood graph $G_{\varepsilon}(V)$ with parameter $\varepsilon$.

	\item Using Definition~\ref{def:VR-expansion} compute
	      $\mathcal{R}_{\varepsilon}$.
\end{enumerate}

This procedure provides us with a geometrical skeleton for the data
cloud points with resolution $\varepsilon$. In case the parameter
$\varepsilon$ is large, there will be more edges connecting points. We
could have a large interconnected graph with little information.
Otherwise, if the parameter $\varepsilon$ is small, there may be fewer
edges connecting the points, resulting in a sparse graph and missing
relevant geometric features within the data cloud.

In the second step we unveil the topological structure of the
neighborhood graph through the Vietoris-Rips complex. The expansion
builds the cofaces related to our simplicial complex. In
Section~\ref{sec:VR-expansion} we will further discuss the algorithm
to achieve this.

\subsection{Neighborhood graph}\label{sec:neigborhood-graph}

The neighborhood graph collects the vertices $V$ and for each vertex
$v\in V$ it adds all the edges $[v, u]$ within the set $u\in V$,
satisfying $d(v,u)\leq\varepsilon$. This brute-force operation works
in $O(n^{2})$ time. We considered a variety of theoretical examples
and it becomes clear that the scale factor in the data set is
relevant. The scale in one variable may differ with the scale of the
output by many orders of magnitude. Thus, proximity is relative to the
scale of the axis on which the information is presented and the
proximity neighborhood may be misjudged.

The Ishigami model presented in Figure~\ref{fig:scales} below shows
how the proximity neighborhood becomes distorted when scale is taken
into consideration.

\begin{figure}
	\centering \includegraphics[width = 0.95\textwidth]{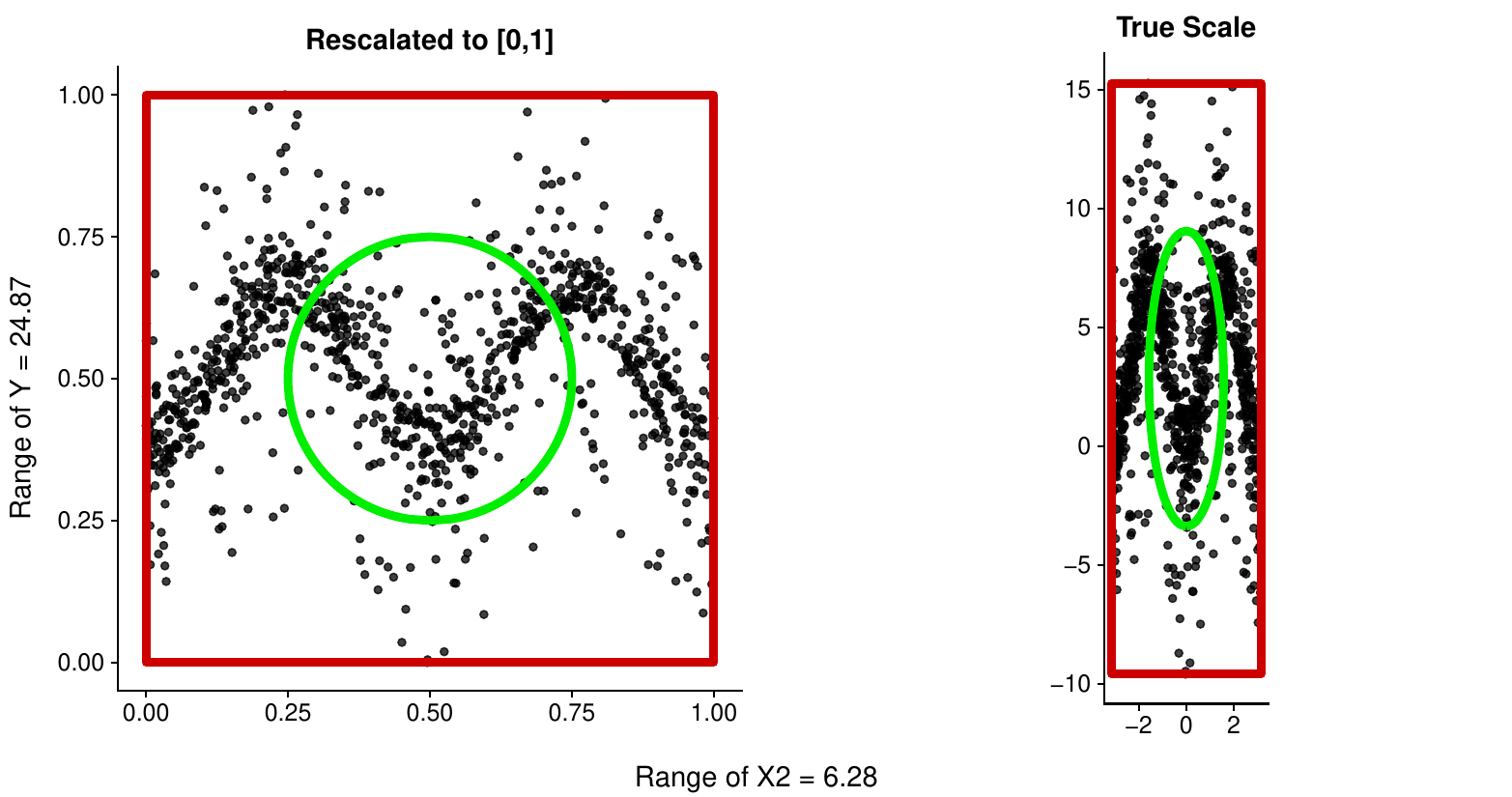}
	\caption{The Second variable of the Ishigami model scaled to $[0,1]$
		with a circle centered at $(0.5, 0.5)$ and radius 1 (left). The same
		circle draw at the true scale of the data (right).}\label{fig:scales}
\end{figure}

We conclude that the aspect ratio between both variables defines how
the algorithm constructs the neighborhood graph. Therefore, to use
circles to build the neighborhood graph we would need to set both
variables to the same scale. Algorithm~\ref{alg:neigborhood-graph}
constructs the VR-neighborhood graph for a cloud of points with
arbitrary scales.

\begin{algorithm}
	\KwData{A set of points $X = (X_{1}, \ldots, X_{n})$ and $Y = (Y_{1}, \ldots, Y_{n})$ \\
	A value $0<\alpha<1$. }%
	\KwResult{The Neighborhood Graph.}

	\SetKwProg{Fn}{Function}{:}{end} \Fn{\textsc{Create-VR-Neighborhood}$((X,Y)$,
	$\alpha$)}{%
	$\displaystyle X_{r} \leftarrow \frac{X - \min{X}}{\max{X}-\min{X}}$ \\
	$\displaystyle Y_{r} \leftarrow \frac{Y - \min{Y}}{\max{Y}-\min{Y}}$ \\
	$n \leftarrow \mathrm{length}(Y)$\\
	\textsf{DistanceMatrix} $\leftarrow$ Matrix$(n\times n)$

	\For{i = 1:n}{%
	\For{j = 1:n}{%
	\textsf{DistanceMatrix}[i,j] $\leftarrow$
	$\sqrt{{(X_{r}\mathrm{[i]}-X_{r}\mathrm{[j]})}^{2} +
		{(Y_{r}\mathrm{[i]}-Y_{r}\mathrm{[j]})}^{2}}$ %
	} %
	} %
	\textsf{$\varepsilon$} $\leftarrow$ \textsc{Quantile} (\textsf{DistanceMatrix}, $\alpha$)\\
	\textsf{AdjacencyMatrix} $\leftarrow$ \textsf{DistanceMatrix} $\leq \varepsilon$ \\

	\textsf{NeighborhoodGraph} $\leftarrow$ \textsc{CreateGraph}
	(\textsf{AdjacencyMatrix}, xCoordinates = $X$, y Coordinates = $Y$)

	\Return (\textsf{NeighborhoodGraph})



	}


	%

	\caption{Procedure to estimate the neighborhood graph given a set of
		points in the plane. } \label{alg:neigborhood-graph}
\end{algorithm}

\begin{enumerate}
	\item Re-scale the points $(X_{i}, Y)$, $i=1,\ldots,n$, onto the
	      square $[0,1]\times[0,1]$.

	\item Estimate the distance matrix between points.

	\item With the distance chart estimate the $\alpha$ quantile of the
	      distances. Declare the radius $\varepsilon_{i}$ as this quantile.

	\item Using Definition~\ref{def:VR-expansion} build the
	      VR-neighborhood graph with $\varepsilon$ changed by $\varepsilon_{i}$
	      for each projection.

\end{enumerate}

\subsection{VR expansion}\label{sec:VR-expansion}

In \cite{Zomorodian2010}  the author describes three methods to build
the Vietoris-Rips complex. The first approach builds the complex by
adding the vertices, the edges and then increasing the dimension to
create triangles, tetrahedrons etc. The second method starts with an
empty complex and adds all the simplices step-by-step stopping at the
desired dimension. In the third method one takes advantage of the fact
that the VR-complex is the combination of cliques in the graph of the
desired dimension.

Due to its simplicity we adopt the third approach and detect the
cliques in the graph. We use the algorithm in~\cite{Eppstein2010}
which is a variant of the classic algorithm from~\cite{Bron1973}. This
algorithm orders the graph $G$ and then computes the cliques using the
Bron-Kerbosch method without pivoting. This procedure reduces the
worst-case scenario from time $\mathcal{O}(3^{n/3})$ to time
$\mathcal{O}(dn3^{d/3})$ where $n$ is the number of vertices and $d$
is the smallest value such that every nonempty sub-graph of $G$
contains a vertex of degree at most $d$.

Constructing a manifold via the VR-complex is not efficient in that
the co-faces may overlap, increasing the computational time. One can
overcome this by creating an ordered neighborhood graph. 

\subsection{Geometrical goodness-of-fit construction}\label{sec:goodness-index}

The main use of the $R^2$ method is to gauge the variability explained by a
chosen model against a null one. In our case, the null model is the
box containing the Vietoris-Rips complex of our data.
This is how we envelop the data in the most basic way possible. Our
model will be the Vietoris-Rips complex itself, since it gives us a
representation of our data through an identifiable structure.

The patterns in the data emerge through the empty spaces in the
projection space generated by each individual variable. When the
point-cloud fills the whole domain  the unknown function
\(\varphi\) applied to \(X_{i}\) produces erratic \(Y\) values.
Otherwise, the function yields a structural pattern which can be
recognized geometrically.


The VR-complex $\mathcal{R}$ estimates the geometric structure of
the data by filling the voids in-between close points. We may then
estimate the area of the created object. This value will not yield too much information about the influence of the variable within the model.
Therefore, we have to  estimate the area of the minimum rectangle
containing the entire object. If some input variable presents a weak
correlation with the output variable, its behavior will be almost
random with uniformly distributed points into the rectangular box. For the other cases, when there is some relevant correlation, it will create a pattern causing empty spaces to appear across the box.


To clarify the notation we will denote by $G_{\varepsilon,i}$ the
neighborhood graph generated by the pair of variables $(X_i,Y)$ and
radius $\varepsilon$. Denote by $\mathcal{R}_{i}$ the
VR-complex generated by $G_{\varepsilon,j}$. We also denote the
geometrical area of the object formed by the VR-complex
$\mathcal{R}_{i}$ by
$\mathrm{Area}(\mathcal{R}_{i})$.

We define the rectangular box for the projection the data $(X_{i},Y)$
as
\begin{equation*}
	B_i=\Bigl[
		\min_{X_i}(\mathcal{R}_{i}),
		\max_{X_i}(\mathcal{R}_{i})\Bigr] \times \Bigl[
		\min_{Y}(\mathcal{R}_{i}),
		\max_{Y}(\mathcal{R}_{i})\Bigr].
\end{equation*}
The geometrical area of $B_i$ will be denoted by $\mathrm{Area}(B_i)$.

Therefore, we can define the measure
\begin{equation*}
	R^2_{\text{Geom},i}
	= 1 - \frac{\mathrm{Area}(\mathcal{R}_{i})}{\mathrm{Area}(B_i)}.
\end{equation*}

\noindent Notice that if the areas of the object and the box are
similar then the index \(  R^2_{\text{Geom},j}\) is close to zero.
Otherwise, if there is a lot of empty space and both areas differ the
index will approach 1.


\section{Results}\label{sec:results}

To measure the quality of the index described above we work concrete
examples. The software used was \textit{R}~cite{RCoreTeam2019}, along
with the packages \texttt{TDA} (\cite{Fasy2019}) for the barcode and
complexes estimation and \texttt{sf} (\cite{Pebesma2018}) for handling
the spatial objects and the area estimation.   An own package
containing all these algorithms will be available soon in CRAN.

\subsection{Theoretical examples}\label{sec:theoretical-examples}

In the following examples we sample $n=1000$ points with the
distribution specified in each case. Due to the number of points in
every example we choose the quantile $5\%$ to determine the radius of
the neighborhood graph. Further insights into this choice will be
presented in the conclusions section.

We will consider five unique settings for our examples, each one with
different topological and geometric features. These settings are not exhaustive and
there are others with interesting features. However through this
sample we do show how the method captures the geometrical correlation
of the variables where other classical methods have failed, as well as
making a case for which method fails to retrieve the desired
information.

The examples considered are the following:

\begin{description}
	\item[Linear:] This is a simple setting with
	      \begin{equation*}
		      Y = 2 X_{1} + X_{2}
	      \end{equation*}
	      and $X_{3}$ is an independent random variable. We set
	      $X_{i}\sim\mathrm{Uniform}(-1,1)$ for $i=1,\ldots, 3$.


	\item[Circle with hole:] The model in this case is
	      \begin{equation*}
		      \begin{cases}
			      X_{1} = r\cos(\theta) \\
			      Y = r\sin(\theta)
		      \end{cases}
	      \end{equation*}
	      with $ \theta \sim \mathrm{Uniform} (0,2\pi)$ and $r \sim \mathrm{Uniform}
		      (0.5,1)$. This form creates a circle with a hole in the middle.

	\item[Multiple circles with holes:] The model consists in three different circles, where we set $ \theta \sim \mathrm{Uniform} (0,2\pi)$:
	      \begin{enumerate}
		      \item Circle centered at $(0,0)$ with radius between 1.5 and 2.5:
		            \begin{equation*}
			            \begin{cases}
				            X_{1} = r_1\cos(\theta) \\
				            Y = r_1\sin(\theta)
			            \end{cases}
		            \end{equation*}
		            where $r_{1}\sim \mathrm{Uniform} (1.5,2.5)$.
		      \item Circle centered at $(3.5,3.5)$ with radius between 0.5 and 1:
		            \begin{equation*}
			            \begin{cases}
				            X_{2} - 3.5 = r_2\cos(\theta) \\
				            Y -3.5 = r_2\sin(\theta)
			            \end{cases}
		            \end{equation*}
		            where $r_{2}\sim \mathrm{Uniform} (0.5,1)$.
		      \item Circle centered at $(-4,4)$ with radius between 1 and 2:
		            \begin{equation*}
			            \begin{cases}
				            X_{3} + 4 = r_3\cos(\theta) \\
				            Y - 4 = r_3\sin(\theta)
			            \end{cases}
		            \end{equation*}
		            where $r_{3}\sim \mathrm{Uniform} (1,2)$.
	      \end{enumerate}

	\item[Ishigami:] The final model is
	      \begin{equation*}
		      Y = \sin X_1 + 7\ \sin^2 X_2 + 0.1\ X_3^4 \sin X_1
	      \end{equation*}
	      where $X_i\sim \mathrm{Uniform}(-\pi, \pi)$ for $i=1,2,3$, $a = 7$ and $b =
		      0.1$.

\end{description}

In order to compare the results with our method, we fit for each case the regression $Y=a_0 + \sum_{k=1}^{10} a_k X_i^k$ for each variable $X_i$. Other regression models  are possible to fit each variable, but we use this as a comparison point to determine how our method performs.


The figures presented in this section represent the estimated manifold
for each input variable $X_{i}$ with respect to the output variable
$Y$. The table below each figure presents the radius used to build the
neighborhood graph, the estimated areas of the manifold object, of the
reference square, and the proposed index.

\begin{figure}
	\centering
	\begin{subfigure}[c]{\textwidth}
		\includegraphics[width=\textwidth]{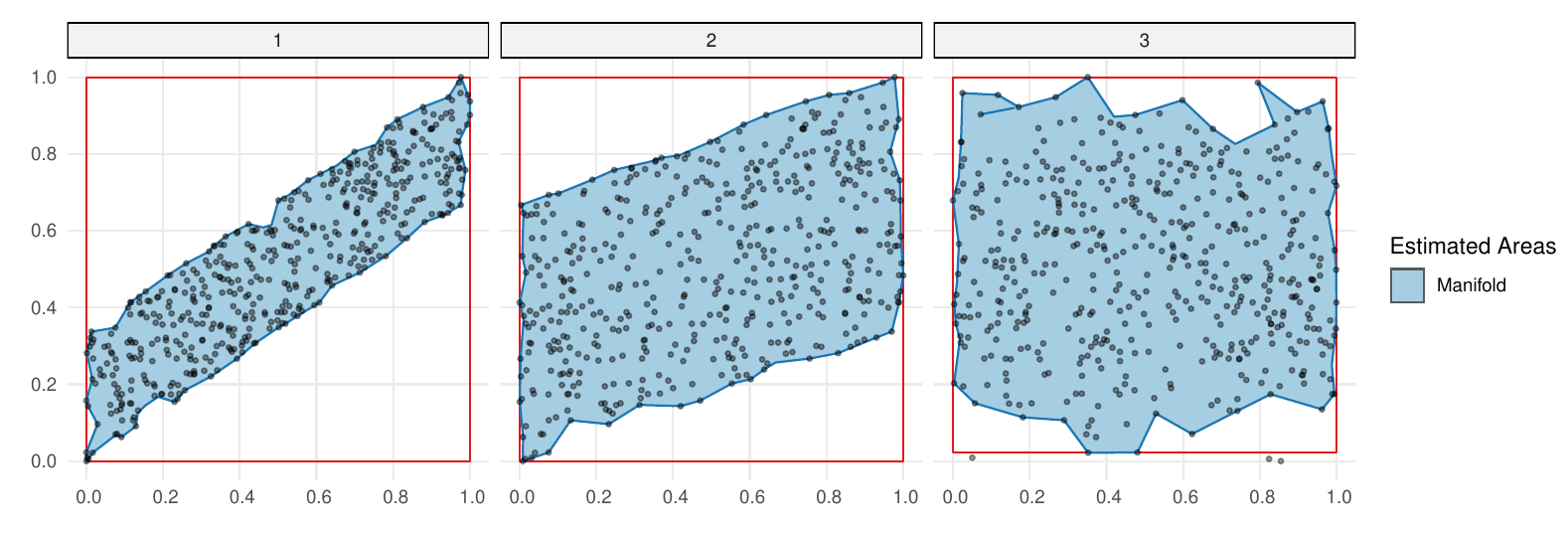}
	\end{subfigure}
	\begin{subtable}{\linewidth}
		\centering
		\begin{tabular}{cccccc}
			\toprule
			Variable
			& $\varepsilon$
			& $\mathrm{Area}(\mathcal{R}_{\varepsilon})$
			& $\mathrm{Area}(B)$
			& $R^2_{\mathrm{Geom}}$
			& $R^2$ \\
			\midrule %
			$X_{1}$ & 0.08 & 0.31 & 1.00 & 0.69 & 0.81 \\
			$X_{2}$ & 0.12 & 0.64 & 1.00 & 0.36 & 0.19 \\
			$X_{3}$ & 0.15 & 0.80 & 0.98 & 0.19 & 0.01 \\
			\bottomrule
		\end{tabular}
	\end{subtable}
	\caption{Results for the linear case.}\label{fig:linear}
\end{figure}

The linear model in Figure~\ref{fig:linear} is simple enough to allow
us to directly see that  variable $X_1$ explains almost  double
the  variability than  variable $X_{2}$. Variables $X_{3}$ to $X_{5}$
will have less important indices. In this case, given the linearity,
the normal $R^2$ score  covers almost 81\% of the variance for the $X_1$. The
rest of the total variance is covered by $X_2$. The rest of variables
have negligible scores. We observe how the empty spaces proliferate
according to the relevance level of the variable. 



\begin{figure}
	\centering
	\begin{subfigure}[b]{\textwidth}
		\includegraphics[width=\textwidth]{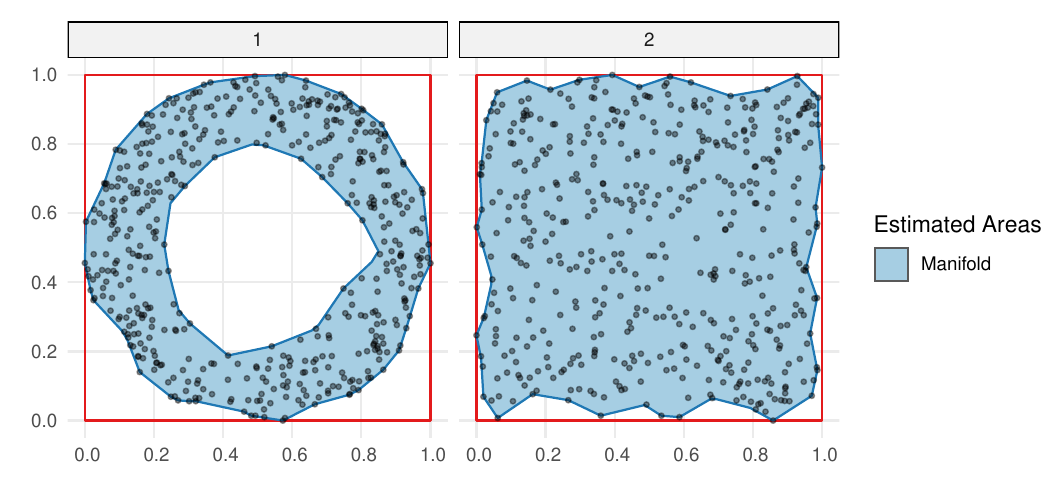}
	\end{subfigure}
	\begin{subtable}{\textwidth}
		\centering
		\begin{tabular}{cccccc}
			\toprule
			Variable
			& $\varepsilon$
			& $\mathrm{Area}(\mathcal{R}_{\varepsilon})$
			& $\mathrm{Area}(B)$
			& $R^2_{\mathrm{Geom}}$
			& $R^2$ \\
			\midrule %
			$X_{1}$ & 0.15 & 0.50 & 1.00 & 0.50 & 0.01 \\
			$X_{2}$ & 0.14 & 0.90 & 1.00 & 0.10 & 0.01 \\
			\bottomrule
		\end{tabular}
	\end{subtable}
	\caption{Results for the circle with 1 hole case.}\label{fig:circle}
\end{figure}

The model of circle with one hole in Figure~\ref{fig:circle} was
discussed in the preliminaries. Recall that in this case both
variables have $R^2$ scores  near  zero for our model even if the geometric
shape showed the contrary. Observe how the first variable has index
equal to $0.46$ and the second one $0.07$. This allows us to say that
the VR-complex built for $X_1$  performs better when explaining the data than just enclosing it in a square. For the second variable the
VR-complex and the enclosed square perform similar, thus the geometric
goodness-of-fit is almost zero.

\begin{figure}
	\centering
	\begin{subfigure}[b]{\textwidth}
		\includegraphics[width=\textwidth]{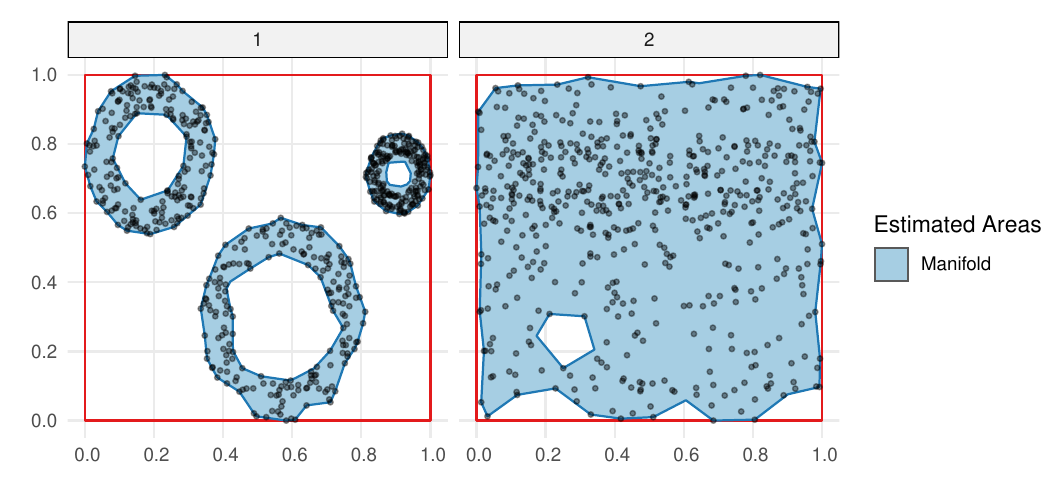}
	\end{subfigure}
	\begin{subtable}{\textwidth}
		\centering
		\begin{tabular}{cccccc}
			\toprule
			Variable
			& $\varepsilon$
			& $\mathrm{Area}(\mathcal{R}_{\varepsilon})$
			& $\mathrm{Area}(B)$
			& $R^2_{\mathrm{Geom}}$
			& $R^2$ \\
			\midrule %
			$X_{1}$ & 0.09 & 0.24 & 1.00 & 0.76 & 0.67 \\
			$X_{2}$ & 0.16 & 0.90 & 1.00 & 0.10 & 0.01 \\
			\bottomrule
		\end{tabular}
	\end{subtable}
	\caption{Results for the circle with 2 holes case.}\label{fig:circle2}
\end{figure}

To test our algorithm further we present the model of connected
circles with holes in Figure~\ref{fig:circle2}. Here we created three
circles with different scales and positions. We could capture the most
relevant features for each projection. In this case the classic $R^2$
could capture only a mere 19\% of the explained variance of the model for the first variable. For the second one the $R^2$ score is near to zero.
Meanwhile, the $R^2_\text{Geom}$ score could detect almost  53\% of
correspondence between the first variable and the outcome. In this
case, our method performs better at detecting the anomalous pattern.

\begin{figure}
	\centering
	\begin{subfigure}[b]{\textwidth}
		\includegraphics[width=\textwidth]{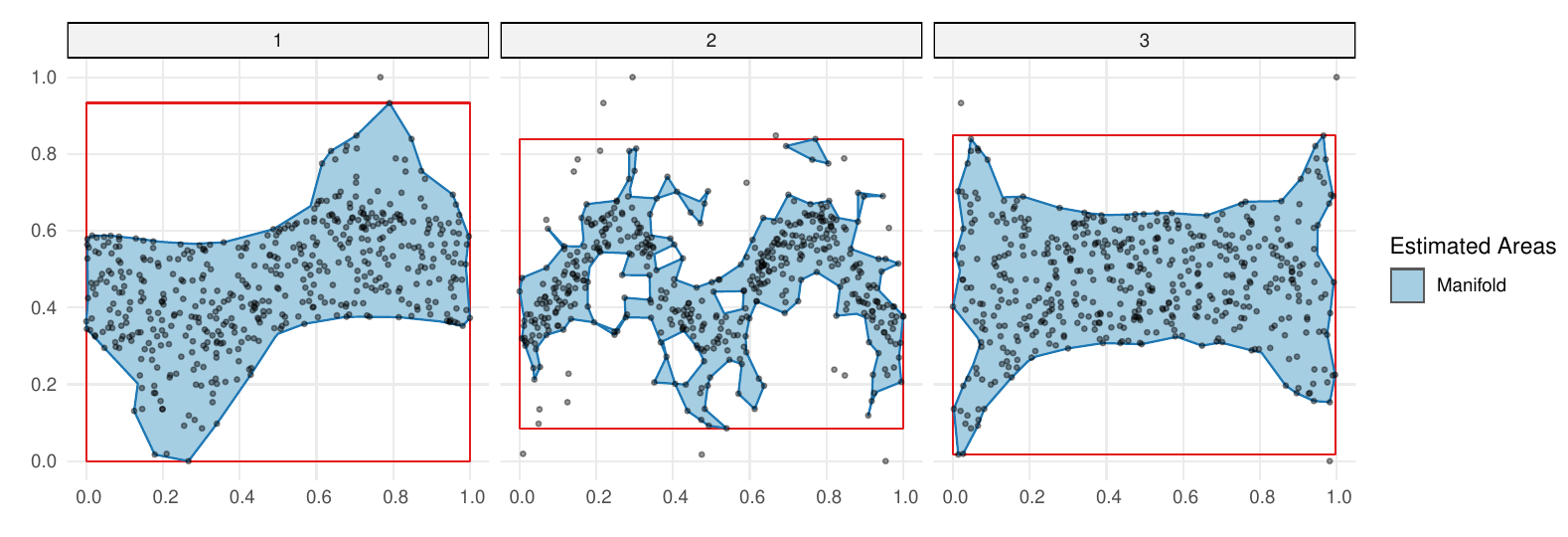}
	\end{subfigure}
	\begin{subtable}{\textwidth}
		\centering
		\begin{tabular}{cccccc}
			\toprule
			Variable
			& $\varepsilon$
			& $\mathrm{Area}(\mathcal{R}_{\varepsilon})$
			& $\mathrm{Area}(B)$
			& $R^2_{\mathrm{Geom}}$
			& $R^2$ \\
			\midrule
			$X_{1}$ & 0.16 & 0.41 & 0.93 & 0.56 & 0.33 \\
			$X_{2}$ & 0.08 & 0.23 & 0.75 & 0.69 & 0.48 \\
			$X_{3}$ & 0.12 & 0.41 & 0.83 & 0.50 & 0.01 \\
			\bottomrule
		\end{tabular}
	\end{subtable}
	\caption{Results for the Ishigami case.}\label{fig:ishigami}
\end{figure}

The final model is produced by the Ishigami function,
Figure~\ref{fig:ishigami}. This is a popular model in sensitivity
analysis because it presents a strong non-linearity and
non-monotonicity with interactions in $X_3$. With other sensitivity
estimators the variables $X_{1}$ and $X_{2}$ have great relevance to
the model, while the third one $X_{3}$ has almost zero. For a further
explanation of this function we refer the reader to~\cite{Sobol1999}.
The first, second  and third variables explain 33\%, 48\% and 1\% of
the variance respectively. In particular, notice how the third
variable is considered pure noise for the regression polynomial model.
Also, notice how in this case, the  $R^2_\text{Geom}$  scores for the three variables are around the 50\% and 60\%. It indicates, for the three variables, that a geometric pattern was revealed by the VR-complex. The three variables present large areas of blank spaces inside their boxes.

\pagebreak

\section{Conclusions and further research}\label{sec:conclusions}

As mentioned above the aim of this paper was to build a goodness-of-fit index relying solely on the topological and geometrical features of a data-cloud. Purely analytic or statistical methods cannot recognize the structure when we project certain variables, primarily when the input is of zero-sum, which might be  artificial noise. In such cases those projections, or the variables in question, have positive conditional variance that might contribute to the model in ways that had not been explored.

Our index proved to be reliable in detecting  variability of the data when the variable is of zero-sum, differentiating between
pure random noise and well-structured inputs. Where the model presents pure noise, our index coincides fully with other methods’ indexes. It also detects relevant structured inputs, in the other cases our index shows the structure in all the variables, which was the notion we wanted to explore.

We do not have yet an efficient and transparent index that measures the geometric goodness-of-fit through the VR-complex. To reach this point we identify a series of research problems to be dealt in the near future: Construct a better and faster to have the base graph. Alternatively, change it to make the process more efficient and hence allow ourselves to run more sophisticated examples, both theoretical and real data examples from controlled examples. One of the central points to be discussed and studied further is determining the radius of proximity, which we believe must
be given by the data-set itself, probably by a more detailed application of persistent homology. Finally, we look forward to
extending our method to more than one variable at a time, to cross-check relevance between variables.

Our major goal for a future project is to asses whether our model is useful in determining the relevance of a variable within a model.

%
%
%

\appendix
\section*{Supplementary Material}

\begin{description}

	\item[R-package for TopSA routine:] R-package “TopSA” estimates
	      sensitivity indices reconstructing the embedding manifold of the
	      data. The reconstruction is done via a Vietoris Rips with a fixed
	      radius. Then the homology of order 2 and the indices are estimated.
	      \url{https://github.com/maikol-solis/topsa}

\end{description}

\bibliographystyle{vancouver} %
\bibliography{sensitivityManifoldsBiblio}







\end{document}